# ACCRETION DISKS

H.C. SPRUIT

*Max-Planck-Institut für Astrophysik, Postfach 1523, D-85740 Garching, Germany*

**Abstract.** In this lecture the basic concepts used in accretion disk theory are reviewed, with emphasis on aspects relevant for X-Ray Binaries and Cataclysmic Variables.

## 1. Introduction

Accretion disks are inferred to exist in objects of very different scales, ranging from the neutron-star-to-solar scale in Low Mass X-Ray Binaries (LMXB) and Cataclysmic Variables (CV), the solar-to-AU scale in protostellar disks, and the AU to parsec scale for the disks in Active Galactic Nuclei (AGN). On the kiloparsec scale galaxies, especially spirals share some properties with accretion disks and can be regarded as the only 'disks' whose structure is, at present, directly observable (the differences are sufficiently large however, that I will exclude them from further discussion in this lecture).

Lacking direct (i.e. spatially resolved) observations of disks, theory has tried to provide models, with varying degrees of success. Since some basic questions about disks are still unsolved (in particular the source of viscosity), the theory of disks is still a fairly undivided field. Progress made by observations or modeling of a particular class of objects is likely to have direct impact for the understanding of other objects.

An interesting connection exists of accretion disks with jets (such as the spectacular jets from AGN and protostars), and outflows (the 'CO-outflows' from protostars and possibly the 'broad-line-regions' in AGN). A magnetically driven wind from the inner regions of an accretion disk is currently to most promising model for these phenomema. See Blandford (1989), Königl and Ruden (1993), Wardle and Königl (1993), Emmering et al. (1992), Spruit (1993).

In this lecture I concentrate on the more basic aspects of accretion disks. Important subjects like theories for the viscosity in disks are dicussed only briefly (section 8), because of complexity and lack of conclusiveness of the current state of the art. Emphasis is on those aspects of accretion disk theory that connect to the observations of LMXB and CV's.

For other reviews on the basics of accretion disks, see Pringle (1981), Verbunt (1982), Frank, King and Raine (1992), Treves et al., (1988).

## 2. Accretion: general

Gas falling into a point mass potential

$$\Phi = -\frac{GM}{r}$$

from a distance $r_0$ to a distance $r$ converts gravitational into kinetic energy, by an amount $GM(1/r - 1/r_0)$ or for simplicity, assuming that the starting distance is large, $GM/r$. If the gas is then brought to rest, for example at the surface of a star, the amount of energy $e$ dissipated per unit mass is

$$e = \frac{GM}{r} \qquad (\text{rest})$$

or, if it goes into a circular orbit at distance $r$:

$$e = \frac{1}{2}\frac{GM}{r} \qquad (\text{orbit}).$$

The dissipated energy goes into internal energy of the gas and into radiation which escapes to infinity (usually in the form of photons, but neutrino losses can also play a role).

### 2.1. ADIABATIC ACCRETION

Consider first the case when radiation losses are neglected. This is *adiabatic* accretion. For an ideal gas with constant ratio of specific heats $\gamma$, the internal energy per unit mass is

$$u = \frac{P}{(\gamma-1)\rho}.$$

With the equation of state

$$P = \mathcal{R}\rho T/\mu \qquad (1)$$

where $\mathcal{R}$ is the gas constant, $\mu$ the mean weight per particle, we find the temperature of the gas after the dissipation has taken place (assuming that the gas goes into a circular orbit):

$$T = \frac{1}{2}(\gamma-1)T_{\text{vir}}, \qquad (2)$$

where $T_{\text{vir}}$, the *virial temperature* is given by

$$T_{\text{vir}} = \frac{GM\mu}{\mathcal{R}r}.$$

In an atmosphere with temperature near $T_{\text{vir}}$, the sound speed is close to the escape speed from the system, the hydrostatic pressure scale height is of the order of $r$, and such an atmosphere evaporates on a relatively short time scale in the form of a stellar wind.

A simple example is *spherical* adiabatic accretion, first discussed by Bondi (1952). The main finding is that such accretion is possible only if $\gamma < 5/3$. This makes sense: the larger $\gamma$, the larger the temperature in the accreted gas (eq. 2), and beyond a critical value the temperature is too high for the gas to stay bound in the potential. A classical situation where adiabatic and roughly spherical accretion takes place is the supernova implosion: when the central temperature becomes high enough for the radiation field to start desintegrating nuclei, $\gamma$ drops and the envelope collapses onto the forming neutron star via a nearly static accretion shock. Another case are Thorne-Żytkov objects (see Ergma, this volume), where $\gamma$ can drop to low values due to pair creation, initiating an adiabatic accretion onto the black hole.

Adiabatic spherical accretion is fast, taking place on the dynamical or free fall time scale

$$\tau_\mathrm{d} = r/v_\mathrm{k} = (r^3/GM)^{1/2}. \tag{3}$$

When radiative loss becomes important, the accreting gas can stay cool irrespective of the value of $\gamma$, and Bondi's critical value $\gamma = 5/3$ plays no role. With losses, the temperatures of accretion disks are usually much lower than the virial temperature. The optical depth of the accreting flow increases with the accretion rate $\dot M$. When the optical depth becomes large enough so that the photons are 'trapped' in the flow, the accretion just carries them in, together with the gas (Rees 1978, Begelman 1979). Adiabatic accretion therefore occurs above a certain critical rate $\dot M_\mathrm{c}$. To compute this rate, consider photons with a mean free path $l$ given by

$$l = (\kappa\rho)^{-1},$$

where $\kappa$ is the opacity (cm$^2$/g) and $\rho$ the density. Over a length scale $r$ these photons diffuse outward at a speed $v_\mathrm{diff}$:

$$v_\mathrm{diff} \sim lc/r.$$

Assuming spherical symmetry the accretion velocity is, in terms of the accretion rate:

$$v_\mathrm{r} = \frac{\dot M}{4\pi r^2 \rho}.$$

The critical rate is obtained when the outward diffusion speed just balances the accretion speed, which yields

$$\dot M_\mathrm{c} = 4\pi rc/\kappa.$$

This happens to be equal to the Eddington rate discussed below. In most cases, $\dot M_\mathrm{c}$ marks the boundary between adiabatic and cooling accretion (for special circumstances where the critical rate is not the Eddington rate, see Spruit et al. 1987). In a fully ionized gas with radiation pressure negligible, $\gamma = 5/3$, so that Bondi's condition for adiabatic accretion is only marginally satisfied. If radiation pressure dominates, $\gamma$ is of the order $4/3$, and adiabatic spherical accretion is possible.

## 2.2. THE EDDINGTON LIMIT

Objects of high luminosity have a tendency to blow their atmospheres away due to the radiative force exerted when the outward traveling photons are scattered or absorbed. Consider a volume of gas on which a flux of photons is incident from one side. Per gram of matter, the gas presents a scattering (or absorbing) surface area of $\kappa$ cm$^2$. The force exerted by the radiative flux $F$ on one gram is $F\kappa/c$. The force of gravity pulling back on this one gram of mass is $GM/r^2$. The critical flux at which the two forces balance is

$$F_{\rm E} = \frac{c}{\kappa}\frac{GM}{r^2} \qquad (4)$$

Assuming that the flux is spherically symmetric, this can be converted into a critical luminosity

$$L_{\rm E} = 4\pi GMc/\kappa, \qquad (5)$$

the Eddington luminosity (e.g. Rybicki and Lightman, 1979). If the gas is fully ionized, its opacity is dominated by electron scattering, and for solar composition $\kappa$ is then of the order 0.3 cm$^2$/g (about a factor 2 lower for fully ionized helium, a factor up to $10^3$ higher for partially ionized gases). With these assumptions,

$$L_{\rm E} \approx 1.7\,10^{38}\frac{M}{M_\odot}\ {\rm erg/s} \approx 4\,10^4\frac{M}{M_\odot}\ L_\odot$$

If this luminosity results from accretion, it corresponds to the Eddington accretion rate $\dot M_{\rm E}$:

$$\frac{GM\dot M_{\rm E}}{r} = L_{\rm E} \quad \rightarrow \quad \dot M_{\rm E} = 4\pi rc/\kappa. \qquad (6)$$

Whereas $L_{\rm E}$ is a true limit that can not be exceeded by a static radiating object except by geometrical factors of order unity (see chapter 10 in Frank et al, 1992), no maximum exists on the accretion rate. For $\dot M > \dot M_{\rm E}$ the plasma is just swallowed whole, including the radiation energy in it (cf. discussion in the preceding section). With $\kappa = 0.3$:

$$\dot M_{\rm E} \approx 1.3\,10^{18}r_6\ {\rm g/s} \approx 2\,10^{-8}r_6\ M_\odot{\rm yr}^{-1},$$

where $r_6$ is the radius of the accreting object in units of 10 km.

## 3. Accretion with Angular Momentum

When the accreting gas has a zonzero angular momentum with respect to the accreting object, it can not accrete directly. A new time scale appears, the time scale for outward transport of angular momentum. Since this is in general much longer than the dynamical time scale, much of what was said about spherical accretion needs modification for accretion with angular momentum.

Consider the accretion in a close binary consisting of a compact (white dwarf, neutron star or black hole) primary of mass $M_1$ and a main sequence companion of mass $M_2$. The mass ratio is defined as $q = M_2/M_1$ (note: $q$ is just as often defined the other way around).

If $M_1$ and $M_2$ orbit each other in a circular orbit and their separation is $a$, the orbital frequency $\Omega$ is

$$\Omega^2 = G(M_1 + M_2)/a^3.$$

The accretion process is most easily described in a coordinate frame that corotates with this orbit, and with its origin in the center of mass. Matter that is stationary in this frame experiences an effective potential, the *Roche potential* (Ch. 4 in Frank, King and Raine, 1992), given by

$$\phi_R(\mathbf{r}) = -\frac{GM}{r_1} - \frac{GM}{r_2} - \frac{1}{2}\Omega^2 r^2 \qquad (7)$$

where $r_{1,2}$ are the distances of point **r** to stars 1,2. Matter that does *not* corotate experiences very different forces (due to the coriolis force). The Roche potential is therefore useful only in a rather limited sense; for non corotating gas intuition based on the Roche geometry is usually confusing. Keeping in mind this limitation, consider the equipotential surfaces of (7). The surfaces of stars $M_{1,2}$, assumed to corotate with the orbit, are equipotential surfaces of (7). Near the centers of mass (at low values of $\phi_R$) they are unaffected by the other star, at higher $\Phi$ they are distorted and at a critical value $\Phi_c$ the two parts of the surface touch. This is the critical Roche surface $S_c$ whose two parts are called the Roche lobes. Binaries lose angular momentum through gravitational radiation and a magnetic wind from the secondary (if it has a convective envelope). Through this loss the separation between the components decreases and both Roche lobes decrease in size. Mass transfer starts when $M_2$ fills its Roche lobe, and is maintained by the angular momentum loss from the system. A stream of gas then flows through the point of contact of the two parts of $S_c$, the inner Lagrange point $L_1$. If the force acting on it were derivable entirely from (7) the gas would just fall in radially onto $M_1$. As soon as it moves however, it does not corotate any more and its orbit under the influence of the coriolis force is different (Fig. 1).

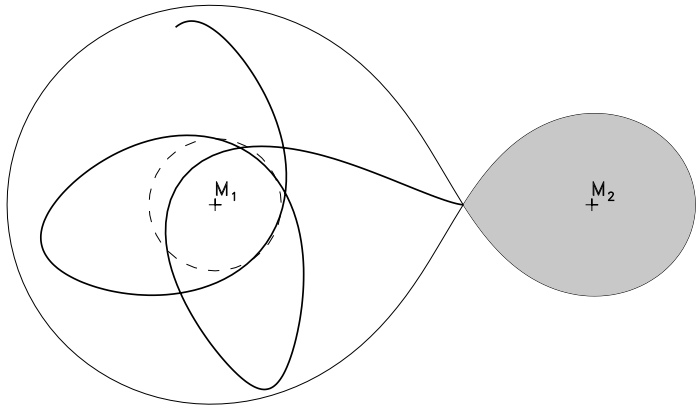

Fig. 1. Roche geometry for $q = 0.2$, with free particle orbit from $L_1$. Dashed: circularization radius.

Since the gas at $L_1$ is very cold compared with the virial temperature, its sound speed is small compared with the velocity it gets after only a small distance from $L_1$.

The flow into the Roche lobe of $M_1$ is therefore highly *supersonic*. Such hypersonic flow is essentially ballistic, that is, the stream flows along the path of free particles.

Though the gas stream on the whole follows an orbit close to that of a free particle, a strong shock develops at the point where the orbit intersects itself. [In practice shocks already develop shortly after passing the pericenter at $M_1$, when the gas is decelerated again. Supersonic flows that are decelerated by whatever means in general develop shocks (e.g. Courant and Friedrichs 1948, Massey, 1968). The effect can be seen in action in the movie published in Różyczka and Spruit, 1993]. After this, the gas settles into a ring, into which the stream continues to feed mass. If the mass ratio $q$ is not too small this ring forms fairly close to $M_1$. An approximate value for its radius is found by noting that near $M_1$ the tidal force due to the secondary is small, so that the angular momentum of the gas with respect to $M_1$ is approximately conserved. If the gas continues to conserve angular momentum while dissipating energy, it settles into the minimum energy orbit with the specific angular momentum $j$ of the incoming stream. The radius of this orbit, the *circularization radius* $r_c$ is determined from

$$(GM_1 r_c)^{1/2} = j.$$

The value of $j$ is found by a simple integration of the orbit starting at $L_1$ and measuring $j$ at some point near pericenter. In units of the orbital separation $a$, $r_c$ and the distance $r_{L1}$ from $M_1$ to $L_1$ are functions of the mass ratio only. As an example for $q = 0.2$, $r_{L1} \approx 0.66a$ and the circularization radius $r_c \approx 0.16a$. In practice the ring forms slightly outside $r_c$ because there is some angular momentum redistribution in the shocks that form at the impact of the stream on the ring.

The evolution of the ring depends critically on nature and strength of the angular momentum transport processes. If sufficient 'viscosity' is present it spreads inward and outward into a disk.

At the point of impact of the stream on the disk the energy dissipated is a significant fraction of the orbital kinetic energy, hence the gas heats up to a significant fraction of the virial temperature. For a typical system with $M_1 = 1M_\odot$, $M_2 = 0.2M_\odot$ having an orbital period of 2 hrs, the observed size of the disk (e.g. Wood et al. 1989b, Rutten et al., 1992) $r_d/a \approx 0.3$, the orbital velocity at $r_d$ about 900 km/s, the virial temperature at $r_d$ is $10^8$K. The actual temperatures at the impact point are much lower, due to rapid cooling of the shocked gas. Nevertheless the impact gives rise to a prominent 'hot spot' in many systems, and an overall heating of the outermost part of the disk.

## 4. Thin disks: properties

### 4.1. Flow in a cool disk is supersonic

The equation of motion, ignoring viscosity, in the potential of a point mass is

$$\frac{\partial \mathbf{v}}{\partial t} + \mathbf{v} \cdot \nabla \mathbf{v} = -\frac{1}{\rho}\nabla P - \frac{GM}{r^2}\hat{\mathbf{r}}, \tag{8}$$

where $\hat{\mathbf{r}}$ is a unit vector in the spherical radial direction $r$. To compare order of magnitudes of the terms, choose a position $r_0$ in the disk, and choose as typical time and velocity scales the orbital time scale $\Omega_0^{-1} = (r_0^3/GM)^{1/2}$ and velocity $\Omega_0 r_0$. Assuming for simplicity an isothermal gas, the pressure gradient term is

$$\frac{1}{\rho}\nabla P = \frac{\mathcal{R}}{\mu}T\nabla \ln \rho.$$

In terms of the dimensionless quantities

$$\tilde{r} = r/r_0, \qquad \tilde{v} = v/(\Omega_0 r_0),$$
$$\tilde{t} = \Omega_o t, \qquad \tilde{\nabla} = r_0 \nabla,$$

the equation of motion is then

$$\frac{\partial \tilde{\mathbf{v}}}{\partial \tilde{t}} + \tilde{\mathbf{v}} \cdot \tilde{\nabla}\tilde{\mathbf{v}} = -\frac{T}{T_{\rm vir}}\tilde{\nabla}\ln \rho - \frac{1}{\tilde{r}^2}\hat{\mathbf{r}}. \qquad (9)$$

All terms and quantities in this equation are of order unity by the assumptions made, except the pressure gradient term which has the coefficient $T/T_{\rm vir}$. If cooling is important, so that $T/T_{\rm vir} \ll 1$, the pressure term is negligible to first approximation, and vice versa. Equivalent statements are also that the gas moves hypersonically on nearly Keplerian orbits, and that the disk is thin, as is shown next.

4.2. DISK THICKNESS

The thickness of the disk is found by considering its equilibrium in the direction perpendicular to the disk plane. In an axisymmetric disk, using cylindrical coordinates $(\varpi, \phi, z)$, measure the forces at a point $\mathbf{r}_0$ $(\varpi, \phi, 0)$ in the midplane, in a frame rotating with the Kepler rate $\Omega_0$ at that point. The gravitational acceleration $-GM/r^2\,\hat{\mathbf{r}}$ balances the centrifugal acceleration $\Omega_0^2 \varpi$ at this point, but not at some distance $z$ above it because gravity and centrifugal acceleration work in different directions. Expanding both accelerations near $\mathbf{r}_0$, one finds a residual acceleration toward the midplane of magnitude

$$g_z = -\Omega_0^2 z.$$

Assuming again an isothermal gas, the condition for equilibrium in the $z$ direction under this acceleration yields a hydrostatic density distribution

$$\rho = \rho_0(\varpi) \exp\left(-\frac{z^2}{2H^2}\right).$$

$H$, the *scale height* of the disk, is given in terms of the isothermal sound speed $c_{\rm s} = (\mathcal{R}/\mu T)^{1/2}$ by

$$H = c_{\rm s}/\Omega_0.$$

We define $\delta \equiv H/r$, the *aspect ratio* of the disk, and find that it can be expressed in several equivalent ways:

$$\delta = \frac{H}{r} = \frac{c_{\rm s}}{\Omega r} = M^{-1} = \left(\frac{T}{T_{\rm vir}}\right)^{1/2},$$

where $M$ is the Mach number of the orbital motion.

4.3. VISCOUS SPREADING

The shear flow between neighboring Kepler orbits in the disk causes friction due to viscosity. The frictional torque is equivalent to exchange of angular momentum between these orbits. But since the orbits are close to Keplerian, a change in angular momentum of a ring of gas also means it must move in position. If the angular momentum is increased, the ring moves to a larger radius. In a thin disk angular momentum transport (more precisely a nonzero divergence of the angular momentum flux) therefore automatically implies redistribution of mass in the disk.

A simple example (Lüst 1952, see also Lynden-Bell and Pringle 1974) is a narrow ring of gas at some distance $r_0$. If at $t = 0$ this ring is released to evolve under the viscous torques, one finds that it first spreads into an asymmetric hump with a long tail to large distances. As $t \to \infty$ one finds that the hump flattens in such a way that almost all the *mass* of the ring is accreted onto the center, while a vanishingly small fraction of the gas carries almost all the angular momentum to infinity. As a result of this asymmetric behavior essentially all the mass of a disk can accrete, even if there is no external torque to remove the angular momentum.

4.4. OBSERVATIONS OF DISK VISCOSITY

Evidence for the strength of the angular momentum transport processes in disks comes from observations of variability time scales. This evidence is not good enough to determine whether the processes really have the same effect as a viscosity, but if this is assumed, estimates can be made of the magnitude of the viscosity.

Cataclysmic Variables give the most detailed information. These are binaries with white dwarf (WD) primaries and (usually) main sequence companions (for reviews see Meyer-Hofmeister and Ritter 1993, Cordova 1993). A subclass of these systems, the Dwarf Novae, show semiregular outbursts. In the currently most developed theory, these outbursts are due to an instability in the disk (Meyer 1990 and references therein). The outbursts are episodes of enhanced mass transfer of the disk onto the primary, involving a significant part of the whole disk. The decay time of the burst is thus a measure of the viscous time scale of the disk (the quantitative details depend on the model, see Cannizzo et al. 1988):

$$t_{\rm visc} = r_{\rm d}^2/\nu,$$

where $r_{\rm d}$ is the size of the disk. With decay times in the order of days, this yields viscosities of the order $10^{15}$ cm$^2$/s, about 14 orders of magnitude above the microscopic viscosity of the gas.

Other evidence comes from the inferred time scale on which disks around protostars disappear, which is of the order of $10^7$ years (Strom et al, 1993).

4.5. $\alpha$-PARAMETRIZATION

The process responsible for such a large viscosity has not been identified yet. Many processes have been proposed, some of which demonstrably work, though often not with an efficiency as high as the observations of CV outbursts seem to indicate. For other processes, such as magnetic turbulence, no theories with sufficient predictive power are available. In order to compare the viscosities in disks under different conditions, one introduces a dimensionless vsicosity $\alpha$:

$$\nu = \alpha \frac{c_s^2}{\Omega}, \tag{10}$$

where $c_s$ is the isothermal sound speed as before. The quantity $\alpha$ was introduced by Shakura and Sunyaev (1973), as a way of parametrizing our ignorance of the angular momentum transport process (their definition is based on a different formula however, and differs by a constant of order unity).

How large can the value of $\alpha$ be, on theoretical grounds? As a simple model, let's assume that the shear flow between Kepler orbits is unstable to the same kind of shear instabilities found for flows in tubes, channels, near walls and in jets. These instabilities occur so ubiquitously that the fluid mechanics community considers them a natural and automatic consequence (e.g. DiPrima and Swinney 1981, p144 2nd paragraph) of a high Reynolds number:

$$\mathrm{Re} = \frac{LV}{\nu}$$

where $L$ and $V$ are characteristic length and velocity scales of the flow. If this number exceeds about 1000 (for some forms of instability much less), instability and turbulence are generally observed. It has been argued (e.g. Zel'dovich 1981) that for this reason hydrodynamic turbulence is the cause of disk viscosity. Let's look at the consequences of this assumption. If due to shear instability an eddy of radial length scale $l$ develops, it will rotate at a rate given by the rate of shear, $\sigma$, in the flow, here

$$\sigma = r\frac{\partial \Omega}{\partial r} = -\frac{3}{2}\Omega.$$

The velocity amplitude of the eddy is $V = \sigma l$, and a field of such eddies produces a turbulent viscosity of the order (leaving out numerical factors of order unity)

$$\nu_{\mathrm{turb}} = l^2 \Omega. \tag{11}$$

In compressible flows, there is maximum to the size of the eddy set by causality considerations. The force that allows an instability to form an overturning eddy is the pressure, which transports information about the flow at the sound speed. The eddies formed by a shear instability can therefore not move faster than $c_s$, hence their size does not exceed $c_s/\sigma \approx H$. At the same time, the largest eddies formed also have the largest contribution to the turbulent viscosity. Thus we should expect that the turbulent viscosity is given by eddies with size of the order $H$:

$$\nu \sim H^2 \Omega,$$

or

$$\alpha \sim 1.$$

Does hydrodynamical turbulence along these lines exist in disks? This is currently not clear; some arguments are discussed briefly in section 8.

## 5. Thin Disks: equations

Consider a thin (= cool, nearly Keplerian, cf. section 4.2) disk, axisymmetric but not stationary. Using cylindrical coordinates $(r, \phi, z)$, (note that we have changed notation from $\varpi$ to $r$ compared with section 4.2) we define the *surface density* $\Sigma$ of the disk as

$$\Sigma = \int_{-\infty}^{\infty} \rho \mathrm{d}z \approx 2H_0\rho_0, \qquad (12)$$

where $\rho_0$, $H_0$ are the density and scaleheight at the midplane. The approximate sign is used to indicate that the coefficient in front of $H$ in the last expression actually depends on details of the vertical structure of the disk. Conservation of mass, in terms of $\Sigma$ is given by

$$\frac{\partial}{\partial t}(r\Sigma) + \frac{\partial}{\partial r}(r\Sigma v_\mathrm{r}) = 0. \qquad (13)$$

(derived by integration the continuity equation over $z$). Since the disk is axisymmetric and nearly Keplerian, the radial equation of motion reduces to

$$v_\phi^2 = GM/r. \qquad (14)$$

By integrating the $\phi$-equation of motion over height and using (12), one gets an equation for the angular momentum balance:

$$\frac{\partial}{\partial t}(r\Sigma\Omega r^2) + \frac{\partial}{\partial r}(r\Sigma v_\mathrm{r}\Omega r^2) = \frac{\partial}{\partial r}(Sr^3\frac{\partial \Omega}{\partial r}), \qquad (15)$$

where $\Omega = v_\phi/r$, and

$$S = \int_{-\infty}^{\infty} \rho\nu \mathrm{d}z \approx \Sigma\nu. \qquad (16)$$

The second approximate equality in (16) holds if $\nu$ can be considered independent of $z$. The mass flux $\dot{M}$ at any point in the disk is given by

$$\dot{M} = -2\pi r\Sigma v_\mathrm{r} = 6\pi r^{1/2}\frac{\partial}{\partial r}(\nu\Sigma r^{1/2}). \qquad (17)$$

The term on the right hand side in (15) is the viscous torque. It is derived most easily with a physical argument (Pringle, 1981)[1].

---

[1] If you prefer a more formal derivation, the fastest way is to consult Landau and Lifshitz (1959) chapter 15 (hereafter LL). Noting that the divergence of the flow vanishes for a thin axisymmetric disk, the viscous stress $\sigma$ becomes (LL eq. 15.3)

$$\sigma_{ik} = \eta\left(\frac{\partial v_i}{\partial x_k} + \frac{\partial v_k}{\partial x_i}\right),$$

where $\eta = \rho\nu$. The viscous force is

$$F_i = \frac{\partial \sigma_{ik}}{\partial x_k} = \frac{\partial \eta}{\partial x_k}\left(\frac{\partial v_i}{\partial x_k} + \frac{\partial v_k}{\partial x_i}\right) + \eta\Delta v_i.$$

The two terms on the right hand side can then be written in cylindrical or spherical coordinates using LL eqs. (15.15-15.18). Finally the viscous torque is computed from the $\phi$-component of the force, and then integrated over $z$.

Assuming that $\nu$ can be taken constant with height (as long as we are not sure what causes the viscosity this is a reasonable simplification) Eqs. (12-15) can be combined into a single equation for $\Sigma$:

$$r\frac{\partial \Sigma}{\partial t} = 3\frac{\partial}{\partial r}[r^{1/2}\frac{\partial}{\partial r}(\nu\Sigma r^{1/2})]. \tag{18}$$

This is the standard form of the *thin disk diffusion equation*. An important conclusion from this equation is: in the thin disk limit, all the physics which determines the time dependent behavior of the disk enters through one quantitity only, the viscosity $\nu$. This is the main attraction if the thin disk approximation.

5.1. STEADY THIN DISKS

In a steady disk ($\partial/\partial t = 0$) the mass flux $\dot M$ is constant through the disk and equal to the accretion rate onto the central object. From (17) we get the surface density distribution:

$$\nu\Sigma = \frac{1}{3\pi}\dot M \left[1 - \beta\left(\frac{r_{\rm i}}{r}\right)^{1/2}\right], \tag{19}$$

where $r_{\rm i}$ is the inner radius of the disk and $\beta$ is a free parameter appearing through the integration constant. It is related to the flux of angular momentum $F_J$ through the disk:

$$F_J = -\dot M \beta \Omega_{\rm i} r_{\rm i}^2, \tag{20}$$

where $\Omega_{\rm i}$ is the Kepler rotation rate at the inner edge of the disk. If the disk accretes onto an object with a rotation rate $\Omega_*$ *less* than $\Omega_{\rm i}$, one finds (Shakura and Sunyaev, 1973, Lynden-Bell and Pringle, 1974) that $\beta = 1$, independent of $\Omega_*$. The angular momentum flux (torque on the accreting star) is inward (spin-up) and equal to the accretion rate times the specific angular momentum at the inner edge of the disk (eq. 20). For stars rotating near their maximum rate ($\Omega_* \approx \Omega_{\rm i}$) and for accretion onto magnetospheres, which can rotate faster than the disk, the situation is different (Popham and Narayan 1991, Paczyński 1991, Bisnovatyi-Kogan 1993, 1994). Accreting magnetospheres, for example, can *spin down* by interaction with the disk, which in that case has a surface density distribution (19) with $\beta < 1$ (see also Spruit and Taam, 1993).

5.2. DISK TEMPERATURE

In this section I assume accretion onto not-too-rapidly rotating objects so that $\beta = 1$. The surface temperature of the disk, which determines how much energy it loses by radiation, is governed primarily by the energy dissipation rate in the disk, which in turn is given by the accretion rate. From the first law of thermodynamics we have

$$\rho T\frac{{\rm d}S}{{\rm d}t} = -{\rm div}{\bf F} + Q_{\rm v}, \tag{21}$$

where $S$ the entropy per unit mass, $\bf F$ the heat flux (including radiation and any form of 'turbulent' heat transport), and $Q_{\rm v}$ the viscous dissipation rate. For changes which happen on time scales longer than the dynamical time $\Omega^{-1}$, the left hand side is small compared with the terms on the right hand side. Integrating over $z$, the divergence turns into a surface term and we get

$$\sigma_{\rm r} T_{\rm s}^4 = \int_0^\infty Q_{\rm v} \, {\rm d}z, \tag{22}$$

where $\sigma_{\rm r}$ is Stefan-Boltzmann's radiation constant $\sigma_{\rm r} = a_{\rm r} c/4$. Thus the energy balance is *local* (for such slow changes): what is generated by viscous dissipation inside the disk at any radius $r$ is also radiated away from the surface at that position. The viscous dissipation rate is equal to $Q_{\rm v} = \sigma_{ij} \partial v_i / \partial x_j$, where $\sigma_{ij}$ is the viscous stress tensor (see footnote in section 5), and this works out to be $Q_{\rm v} = 9/4\, \Omega^2 \nu \rho$. Eq. (22), using (19) then gives the surface temperature in terms of the accretion rate:

$$\sigma_{\rm r} T_{\rm s}^4 = \frac{9}{8} \Omega^2 \nu \Sigma = \frac{GM}{r^3} \frac{3\dot M}{8\pi} \left[ 1 - \left(\frac{r_{\rm i}}{r}\right)^{1/2} \right]. \tag{23}$$

This shows that the surface temperature of a steady disk, at a given distance $r$ from the accreter, depends *only* on the product $M\dot M$, but not on the highly uncertain value of the viscosity. For $r \gg r_{\rm i}$ we have

$$T_{\rm s} \sim r^{-3/4}. \tag{24}$$

These considerations only tells us about the surface temperature. The internal temperature in the disk is quite different, and depends on the mechanism transporting energy to the surface. Because it is the internal temperature that determines the disk thickness $H$ (and probably also the viscosity), this transport needs to be considered in some detail for realistic disk models. This involves a calculation of the vertical structure of the disk. Because of the local (in $r$) nature of the balance between dissipation and energy loss, such calculations can be done as a grid of models in $r$, without having to deal with exchange of energy between neighboring models. Schemes borrowed from stellar structure computations are used (e.g. Meyer and Meyer-Hofmeister 1982, Pringle et al. 1986, Cannizzo et al. 1988).

An approximation to the temperature in the disk can be found when a number of additional assumption is made. As in stellar interiors radiative transport dominates at high temperatures. The internal temperature is then found from the Eddington approximation for a radiative atmosphere:

$$\sigma_{\rm r} T^4 = \frac{3}{4} \tau F_{\rm r}, \qquad (\tau \gg 1) \tag{25}$$

where $\tau$ is the optical depth and $F_{\rm r}$ the heat flux. Assuming that most of heat is generated near the midplane (which is the case if $\nu$ is constant with height), $F_{\rm r}$ is constant with height and equal to $\sigma_{\rm r} T_{\rm s}^4$. Finally we assume that the opacity $\kappa$ is constant with $z$. The optical depth is then $\tau = \kappa \Sigma / 2$. With (23) this yields for the temperature at the midplane

$$T^4 = \frac{27}{64} \sigma_{\rm r}^{-1} \Omega^2 \nu \Sigma \kappa \tag{26}$$

Assuming equation of state (1), valid when radiation pressure is small, we find with (19) for the disk thickness:

$$\begin{aligned}\frac{H}{r} &= (\mathcal{R}/\mu)^{2/5} \left(\frac{3}{64\pi^2 \sigma_{\rm r}}\right)^{1/10} \alpha^{-1/10} \kappa^{1/10} (GM)^{-7/20} r^{1/20} (f\dot{M})^{1/5} \\ &= 5\, 10^{-3} \alpha^{-1/10} \kappa^{1/10} r_6^{1/20} \left(\frac{M}{M_\odot}\right)^{-7/20} (f\dot{M}_{16})^{1/5}, \qquad (P_{\rm r} \ll P) \end{aligned} \qquad (27)$$

where $r_6 = r/(10^6 \text{ cm})$, $\dot{M}_{16} = \dot{M}/(10^{16} \text{g/s})$, and

$$f = 1 - (r_{\rm i}/r)^{1/2}.$$

From this we conclude that: i) the disk is thin, $H/r < 10^{-2}$ in X-Ray Binaries, ii) the disk thickness is relatively insensitive to the parameters, especially $\alpha$, $\kappa$ and $r$. It must be stressed however that this depends fairly strongly on the assumption that the energy is dissipated in the disk interior. If the dissipation takes place close to the surface [such as in some magnetic reconnection models (Field and Rogers 1993 and references therein)], the disk temperature will be much closer to the surface temperature and $H$ would be even smaller in such disks.

### 5.3. RADIATION PRESSURE DOMINATED DISKS

In the inner regions of disks in XRB, the radiation pressure can dominate over the gas pressure, which results in a different expression for the disk thickness. The total pressure $P$ is

$$P = P_{\rm r} + P_g = \frac{1}{3} a T^4 + P_g. \qquad (28)$$

The isothermal sound speed is given by $c_{\rm s}^2 = P/\rho$, and the relation $c_{\rm s} = \Omega H$ still holds. For $P_{\rm r} \gg P_g$ we get from (25), with (23)

$$cH^2 = \frac{3}{16\pi} \frac{\Sigma}{\rho_0} \kappa f \dot{M}.$$

With the rather approximate relation $\Sigma = 2H\rho_0$ we get

$$\frac{H}{R_*} \approx \frac{3}{8\pi} \frac{\kappa}{cR_*} f\dot{M} = \frac{3}{2} f \frac{\dot{M}}{\dot{M}_{\rm E}}, \qquad (29)$$

where $R_*$ is the stellar radius and $\dot{M}_{\rm E}$ the Eddington rate for this radius. It follows that the disk becomes thick near the star, if the accretion rate is near Eddington (though this is mitigated somewhat by the decrease of the factor $f$). Accretion near the Eddington limit is geometrically not thin any more, and in addition other processes such as angular momentum loss by 'photon drag' (Lamb, this volume) have to be taken into account (see also Miller, this volume).

## 6. Comparison with CV observations

The number of meaningful quantitative tests between theory of disks and observations is somewhat limited since in the absence of a theory for $\nu$, it is a bit meagre on predictive power. The most detailed information perhaps comes from modeling of CV outbursts.

Two simple tests are possible (nearly) independently of $\nu$. These are the prediction that the disk is geometrically quite thin (eq. 27) and the prediction that the surface temperature $T_s \sim r^{-3/4}$ in a steady disk. The latter can be tested in a subclass of the CV's that do not show outbursts, the nova-like systems, which are believed to be approximately steady accreters. If the system is also eclipsing, eclipse mapping techniques can be used to derive the brightness distribution with $r$ in the disk (Horne, 1985, 1993). If this is done in a number of colors so that bolometric corrections can be made, the results (e.g. Rutten et al. 1992) show in general a *fair* agreement with the $r^{-3/4}$ prediction. Two deviations occur: i) a few systems show significantly flatter distributions than predicted, and ii) most systems show a 'hump' near the outer edge of the disk. The latter deviation is easily explained, since we have not taken into account that the impact of the stream heats the outer edge of the disk. Though not important for the total light from the disk, it is an important local contribution near the edge.

Eclipse mapping of Dwarf Novae in quiescence give a quite different picture. Here, the inferred surface temperature is often nearly flat (e.g. Wood et al. 1989a). This is understandable however since in quiescence the mass flux depends strongly on $r$. In the inner parts of the disk it is small, near the outer edge it is close to its average value. With eq. (23), this yields a flatter $T_s(r)$. The lack of light from the inner disk is compensated during the outburst, when the accretion rate in the inner disk is higher than average (see Mineshige and Wood 1989 for a more detailed comparison). The effect is also seen in the 2-dimensional hydrodynamic simulations of accretion in a binary by Różyczka and Spruit (1993). These simulations show an outburst during which the accretion in the inner disk is enhanced, between two episodes in which mass accumulates in the outer disk. The bolometric surface brightness in this simulation is shown in Fig. 2.

## 7. Comparison with LMXB observations

In low mass X-ray binaries a complication arises because of the much higher luminosity of the accreting object. Since a neutron star is roughly 100 times smaller than a white dwarf, it produces 100 times more luminosity for a given accretion rate.

Irradiation of the disk by the central source leads to a different surface temperature than predicted by (23). The central source (star plus inner disk) radiates the total accretion luminosity $GM\dot{M}/R_*$ (assuming sub-Eddington accretion, see section 2). If the disk is *concave*, it will intercept some of this luminosity. If the central source is approximated as a point source, and the angle between the disk surface and the radial direction is $\epsilon$, the irradiating flux on the disk surface is

$$F_{\rm irr} = \epsilon \frac{GM\dot{M}}{4\pi R_* r^2}, \qquad \epsilon = {\rm d}H/{\rm d}r - H/r, \tag{30}$$

or

$$\frac{F_{\rm irr}}{F_{\rm d}} = \frac{1}{3}\frac{\epsilon}{f}\frac{r}{R_*},$$

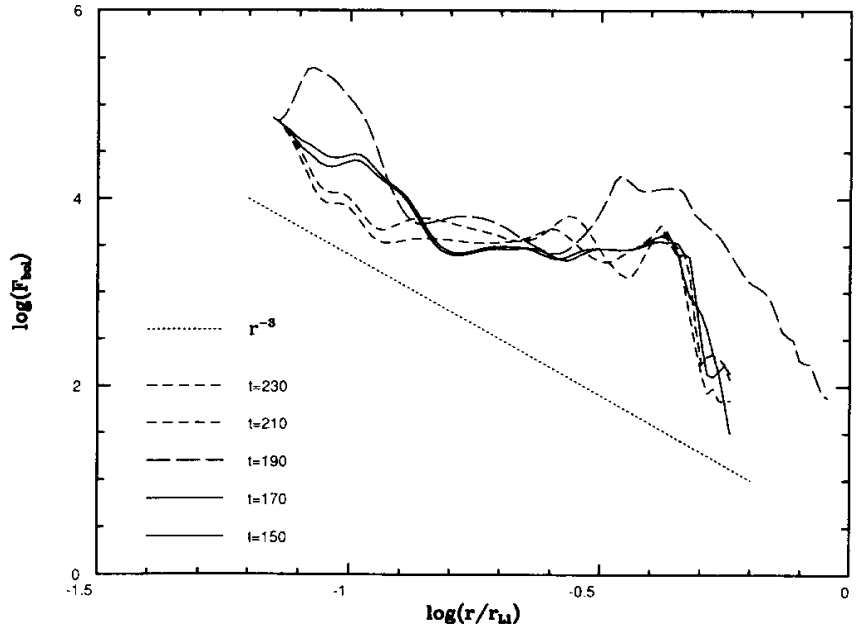

Fig. 2.  Bolometric surface brightness in numerical simulations of Różyckzka and Spruit, 1993. Long dashed: during, solid: before, short dashed: after the outburst. Note hump at the disk edge, due to the impact of the stream.

where $F_d$ is the flux generated internally in the disk, given by (23). On average, the angle $\epsilon$ is of the order of the aspect ratio $\delta$. With $f \approx 1$, and our fiducial value $\delta \approx 5\,10^{-3}$, we find that irradiation dominates for $r > 10^9$cm. This is compatible with observations (for reviews see Mason 1990, Van Paradijs and McClintock 1993), which show that the optical and UV are completely dominated by reprocessed radiation. When irradiation by an external source is included in the thin disk model, the energy balance equation (23) becomes

$$\sigma_r T_s^4 = \frac{GM}{r^3}\frac{3\dot{M}}{8\pi}f + (1-a)F_{\rm irr}, \tag{31}$$

where $a$ is the X-ray albedo of the surface, i.e. $1-a$ is the fraction of the incident flux that is absorbed in the *optically thick* layers of the disk (photons absorbed higher up only serve to heat up the corona of the disk). In the region where the irradiating flux dominates the first term on the rhs can be neglected in computing the surface temperature. For $F_{\rm irr}$ we need the disk thickness $H$. It turns out that the irradiating flux, even when it is larger than the internally generated heat flux, has only a small effect on the disk thickness, as long as $F_{\rm irr}/F_d < \tau$, where $\tau$ is the optical thickness of the disk (Lyutyi and Sunyaev 1976, Vrtilek et al. 1991). Here, we are assuming again that the energy is transported vertically by radiation. The reason for this is the same as in radiative envelopes of stars, which are also insensitive to the surface boundary condition. In the reprocessing region of the disks of LMXB, the conditions are such that $F_{\rm d} \ll F_{\rm irr} < \tau F_{\rm d}$, hence we must use eq. (27) for $H$. This yields $\epsilon = (21/20)H/r \approx 5\,10^{-3}$, and $T_{\rm s} \sim r^{0.5}$, and we still expect the disk to remain thin.

From the paucity of sources in which the central source is eclipsed by the companion one deduces that the companion is barely or not at all visible from the inner disk, presumably because the outer parts of the disk are much thicker than expected from the above arguments. This is consistent with the observation that the characteristic modulation of the optical light curve due to irradiation of the secondary's surface by the X-rays is not very strong in LMXB (with the exception of Her X-1, which has a large companion). The place of the eclipsing systems is taken by the so-called 'Accretion Disk Corona' (ADC) systems, where shallow eclipses of a usually extended X-ray source are seen (for reviews of the observations, see Lewin et al. 1995). The conclusion is that there is an extended X-ray scattering 'corona' above the disk. It scatters a few per cent of the X-ray luminosity.

What causes this corona and the large inferred thickness of the disk is not clear at present. The thickness expected from disk theory is a rather stable number. To 'suspend' matter at the inferred height of the disk forces are needed that are much larger than the pressure forces available in an optically thick disk. A thermally driven wind, produced by X-ray heating of the disk surface, has been invoked (Begelman et al., 1983, see also Horn and Meyer, 1993). For other explanations, see van Paradijs and McClintock (1995). Perhaps a magnetically driven wind from the disk, such as seen in protostellar objects (e.g. Königl and Ruden 1993) can explain both the shielding of the companion and the scattering. Such a model would resemble magnetically driven wind models for the broad-line region in AGN (e.g. Emmering et al., 1992).

### 7.1. TRANSIENTS

Soft X-ray transients (also called X-ray Novae) are believed to be binaries similar to the other LMXB, but somehow the accretion is episodic, with very large outbursts recurring on time scales of decades (sometimes years). There are many black hole candidates among these transients (see van Paradijs and McClintock 1995 for a review). As with the Dwarf Novae, the time dependence of the accretion in transients can in principle be exploited to derive information on the disk viscosity, assuming that the outburst is caused by an instability in the disk. The closest relatives of soft transients among the White Dwarf plus main sequence star systems are probably the WZ Sge stars (van Paradijs and Verbunt 1984, Bailyn 1992), which show (in the optical) similar outbursts with similar recurrence times (cf. Warner 1987, O'Donoghue et al. 1991). Like the soft transients, they have low mass ratios ($q < 0.1$). For a given angular momentum loss, systems with low mass ratios have low mass transfer rates, so the speculation is that the peculiar behavior of these systems is somehow connected with a low mean accretion rate.

### 7.2. DISK INSTABILITY

The most developed model for outbursts is the disk instability model of Osaki (1974), Hōshi (1979), Smak (1984), Meyer and Meyer-Hofmeister (1981), Faulkner et al. (1983) (for recent discussions see King 1995, Osaki 1993). In this model the instablity that gives rise to cyclic accretion is due to a temperature dependence of the viscous stress. In any local process that causes an effective viscosity, the resulting $\alpha$- parameter will be a function of the main dimensionless parameter of the disk, the aspect ratio $H/r$. If this is a sufficiently rapidly increasing function, such that $\alpha$ is large in hot disks and low in cool disks, an instability results by the following mechanism. Suppose we start the disk in a stationary state at the mean accretion rate. If this state is perturbed by a small temperature increase, $\alpha$ goes up, and by the increased viscous stress the mass flux $\dot M$ increases. By (23) this increases the disk temperature further, resulting in a runaway to a hot state. Since $\dot M$ is larger than the average, the disk empties partly, reducing the surface density and the central temperature (eq. 26). A cooling front then transforms the disk to a cool state with an accretion rate below the mean. The disk in this model switches back and forth between hot and cool states. By adjusting $\alpha$ in the hot and cool states, or by adjusting the functional dependence of $\alpha$ on $H/r$, outbursts are obtained that agree reasonably with the observations of soft transients (Lin and Taam 1984, Mineshige and Wheeler, 1989). A rather strong increase of $\alpha$ with $H/r$ is needed to get the observed long recurrence times.

Another possible mechanism for instability has been found in 2-D numerical simulations of accretion disks (Blaes and Hawley 1988, Różyczka and Spruit 1993). The outer edge of a disk is found, in these simulations, to become dynamically unstable to a oscillation which grows into a strong excentric perturbation (a crescent shaped density enhancement which rotates at the local orbital period). Shock waves generated by this perturbation spread mass over most of the Roche lobe, at the same time the accretion rate onto the central object is strongly enhanced. This process is different from the Osaki and Hōshi mechanism, since it requires 2 dimensions, and does not depend on the viscosity (instead, the internal dynamics in this instability *generates* the effective viscosity that causes a burst of accretion).

### 7.3. OTHER INSTABILITIES

Instability to heating/cooling of the disk can be the due to several effects. The cooling rate of the disk, if it depends on temperature in an appropriate way, can cause a thermal instability like that in the interstellar medium. Other instabilities may result from the dependence of viscosity on conditions in the disk. For a general treatment see Piran (1978), for a shorter discussion see Treves et al., 1988.

## 8. Sources of Viscosity

The high Reynolds number of the flow in accretion disks (of the order $10^{11}$ in the outer parts of a CV disk) would, to most fluid dynamicists, seem an amply sufficient condition for the occurrence of hydrodynamic turbulence. A theoretical argument against such turbulence often used in astrophysics (Kippenhahn and Thomas 1981, Pringle 1981) is that in cool disks the gas moves almost on Kepler orbits, which are quite stable (except the orbits that get close to the companion). This stability is related to the known stabilizing effect that rotation has on hydrodynamical turbulence (Bradshaw 1969, for a recent discussion see Tritton 1992). Kippenhahn and Thomas also point out that the one laboratory experiment that comes close to the situation in accretion disks, namely the rotating Couette flow, does not become unstable for parameters like in disks (for the rather limited range in Reynolds numbers available).

A (not very strong) observational argument is that hydrodynamical turbulence as described above would produce an $\alpha$ that does not depend on the nature of the disk, so that all objects should have the same value. This is unlikely to be the case. From the modeling of CV outbursts one knows, for example, that $\alpha$ probably increases with temperature (more accurately, with $H/r$, see previous section). Also, there are indications from the inferred life times and sizes of protostellar disks (Strom et al. 1993) that $\alpha$ may be rather small there, $\sim 10^{-3}$, whereas in outbursts of CV's one infers values of the order $0.1 - 1$.

The indeterminate status of the hydrodynamic turbulence issue is one of the most annoying problems in disk theory. Numerical simulations may be of some help, even though Reynolds numbers as high as those in disks can not be reached. The astrophysical approach has been to circumvent the problem by finding plausible alternative mechanisms that might work just as well.

Among the processes that have been proposed repeatedly as sources of viscosity is convection due to a vertical entropy gradient (for recent work see Kley et al. 1993), which may have some limited effect in convective parts of disks. Another class are *waves* of various kinds. Their effect can be global, that is, not reducible to a local viscous term because by traveling across the disk they can communicate torques over large distances. For example, waves set up at the outer edge of the disk by tidal forces can travel inward and by dissipating there can effectively transport angular momentum *outward* (e.g. Narayan et al. 1987, Spruit et al. 1987). A nonlinear version of this idea are selfsimilar spiral shocks, observed in numerical simulations (Sawada et al. 1987) and studied analytically (Spruit 1987). Such shocks can produce accretion at an effective $\alpha$ of 0.01 in hot disks, but are probably not very effective in disks as cool as those in CV's and XRB. A second non-local mechanism is provided by a magnetically accelerated *wind* originating from the disk surface (Blandford 1976, Bisnovatyi-Kogan and Ruzmaikin 1976, Lovelace 1976, Blandford and Payne 1982, for recent reviews see the papers in Lynden-Bell 1993). In principle, such winds can take care of *all* the angular momentum loss needed to make accretion possible in the absence of a viscosity (Königl, 1989). The attraction of this idea is that magnetic winds are a strong contender for explaining the strong outflows and jets seen in protostellar objects and AGN. It is not yet clear however if, even in these objects, the wind is actually the main source of angular momentum loss. In sufficiently cool or massive disks, selfgravitating instabilies of the disk matter can produce internal friction. Paczýnski (1978) has proposed that the resulting heating would limit the instability and keep the disk in a well defined moderately unstable state. The angular momentum transport in such a disk has been modeled by several authors (e.g. Lin and Pringle, 1990). Disks in XRB are always too hot for selfgravity to play a role.

Magnetic forces can be very effective at transporting angular momentum. If it can be shown that the shear flow in the disk produces some kind of small scale fast dynamo process, that is, some form of magnetic turbulence, a large effective $\alpha$ may be expected. An elegant demonstration that disks do indeed generate such fields on a dynamical time scale has recently (Balbus and Hawley 1991, 1992) come with the revival of an old instability meachanism in magnetic shear flows (Velikhov 1959, Chandrasekhar 1961). This instability rapidly stretches a weak initial field into small loops in the $r-z$ plane, which are then stretched into an azimuthal field by the shear. The importance of the mechanism is that it will amplify a weak field even if it has initially no component that can be stretched by the shear flow. The nonlinear development of this instability is almost certainly highly 3-dimensional, involving a mostly small-scale magnetic field. It is not clear that the instability still plays a significant role in this highly dynamic situation; the net effect could well be the magnetic turbulence whose properties have been guessed previously by many authors (e.g. Eardley and Lightman 1975, Pudritz 1981, Meyer and Meyer-Hofmeister 1982). The main difficulty with these models has been the lack of reliability of the end result, the effective $\alpha$. It is conceivable that by 3-D numerical magnetohydrodynamic simulations one may soon get sufficient insight to derive useful numbers.

## 9. Beyond thin disks

Ultimately, much of the progress in developing useful models of accretion disks will depend on detailed numerical simulations in 2 or 3 dimensions. In the disks one is interested in, there is usually a large range in length scales (in LMXB disks, from less than the 10km neutron star radius to the more than $10^5$km orbital scale). Correspondingly, there is a large range in time scales that have to be followed. This not technically possible at present and in the foreseeable future. In numerical simulations one is therefore limited to studying in an approximate way aspects that are either local or of limited dynamic range in $r, t$ (for examples, see Hawley 1991, Różyczka and Spruit 1993). For this reason, there is still a need for approaches that relax the strict thin disk framework somewhat without resorting to full simulations.

### 9.1. 'Slim' disks

Due to the thin disk assumptions, the pressure gradient does not contribute to the support in the radial direction and the transport of heat in the radial direction is negligible. Some of the physics of thick disks can be included in a fairly consistent way in the so-called 'slim disk' approximation (Abramowicz et al., 1988). It has the advantage of remaining one-dimensional, but allows the inclusion of the radial pressure gradient (so that for example sound waves can occur) and radial heat transport. It still ignores all multidimensional processes, but allows one to study those aspects of the physics of thick disks that are approximately one-dimensional.

### 9.2. Boundary layers

In order to accrete onto the star, the disk matter must dissipate an amount of energy given by

$$\frac{GM\dot M}{2R_*}(1-\Omega_*/\Omega_{\rm k}(R_*))^2 \ . \tag{32}$$

The factor in brackets measures the kinetic energy of the matter at the inner edge of the disk $(r=R_*)$, in the frame of the stellar surface. Due to this heating the disk inflates into a 'belt' at the equator of the star, of thickness $H$ and radial extent of the same order. Equating the radiation emitted from the surface of this belt to (32) the surface temperature $T_{\rm sB}$ of the belt, assuming optically thick conditions and a slowly rotating star ($\Omega_*/\Omega_k \ll 1$) is given by

$$\frac{GM\dot{M}}{8\pi R_*^2 H} = \sigma T_{\text{sB}}^4 \tag{33}$$

To find the temperature inside the belt and its thickness, use eq. (25). The value of the surface temperature is higher, by a factor of the order $(R_*/H)^{1/4}$, than the simplest thin disk estimate (23, ignoring the $(r/r_i)^{1/2}$ factor). In practice, this works out to a factor of a few. The surface of the belt is therefore not very hot. The situation is quite different if the boundary layer is not optically thick (Pringle and Savonije 1979). It then heats up to much higher temperatures. Analytical methods to obtain the boundary layer structure have been used by Regev and Hougerat (1988), numerical solutions of the slim disk type by Narayan and Popham (1993), 2-D numerical simulations by Kley (1991). These considerations are primarily relevant for CV disks; in accreting neutron stars, the dominant effects of radiation pressure have to be included.